# Evolution of flexible industrial assembly

Ali Ahmad Malik
School of Engineering & Computer Science
Oakland University, Michigan, United States
*Email: aliahmadmalik@oakland.edu*

**Pre-print**

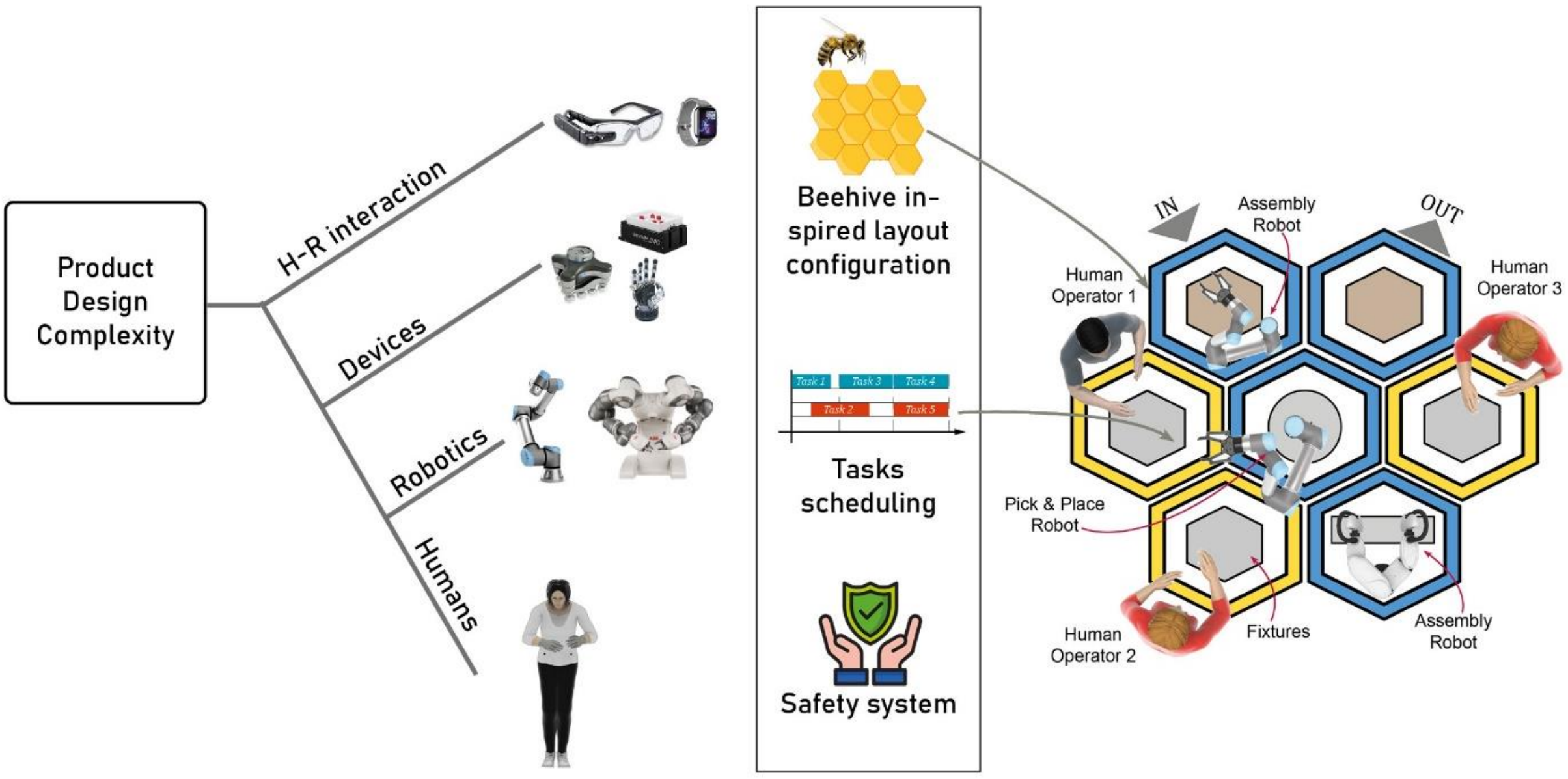

**Highlights**

1. Due to the complexity and variety of assembly tasks, rationalization of manufacturing automation has considerably remained away from assembly systems.
2. With smart manufacturing technologies such as collaborative robots, additive manufacturing, and digital twins, the opportunities have arisen for the next reshaping of assembly systems.
3. This may result into a new manufacturing paradigm driven by the advancement of new technologies, new customer expectations and by establishing new kinds of manufacturing systems.
4. The article presents a beehives inspired framework to develop intelligent reconfigurable and repurposable adaptive assembly (IRRAA) cells.

# Evolution of flexible industrial assembly

Ali Ahmad Malik
School of Engineering & Computer Science
Oakland University, Michigan, United States
*Email: aliahmadmalik@oakland.edu*

**Abstract**

Assembly is a key industrial process to achieve finished goods. Driven by market demographics and technological advancements, industrial assembly has evolved through several phases i.e. craftmanship, bench assembly, assembly lines and flexible assembly cells. Due to the complexity and variety of assembly tasks, besides significant advancement of automation technologies in other manufacturing activities, humans are still considered vital for assembly operations. The rationalization of manufacturing automation has considerably remained away from assembly systems. The advancement in assembly has only been in terms of better scheduling of work tasks and avoiding of wastes. With smart manufacturing technologies such as collaborative robots, additive manufacturing, and digital twins, the opportunities have arisen for the next reshaping of assembly systems. The new paradigm promises a higher degree of automation yet remaining flexible. This may result into a new manufacturing paradigm driven by the advancement of new technologies, new customer expectations and by establishing new kinds of manufacturing systems. This study explores the future collaborative assembly cells, presents a generic framework to develop them and the basic building blocks.

## 1. Introduction

Future factories are believed to be intelligent, reconfigurable and adaptable to market dynamics (Bilberg and Malik, 2019). The vision of batch-size-one production and mass-personalization is not so far from becoming a reality (Jardim-Goncalves, Romero and Grilo, 2017). Technologies are getting available and are becoming *'smart'* with each passing day. Ever-increasing computing power, big data, and smart robots are forming this new wave of smart manufacturing (Lu, Xu and Wang, 2020).

In the pre-industrial era, assembly was carried out as craft-ship (Nof, 2009) . With industrialization, the new form of industrial assembly was established in the form of benchwork (Leviton and William, 1973). Henry Ford revolutionized industrial assembly through assembly line and the societal impact is often referred to as the 2nd industrial revolution (Black, 2007). In the later years, the pursuit to flexibility and effectiveness evolved assembly systems as assembly cells (Chiarini, 2012). These cells can be in many configurations such as U-shaped, L-shaped or S-shaped (Leng *et al.*, 2021). The challenge still continuing with assembly is its high dependence on humans thus sacrificing the opportunities to achieve cost reduction, reduced errors, and relieve humans from repetitive tasks (Malik and Brem, 2021a).

Assembly operations are characterized with a large variety of simple to complex physical tasks requiring human dexterity and flexibility (Weidner, Kong and Wulfsberg, 2013). Many of the tasks are repetitive. The tasks in an assembly process can be physical or cognitive (Romero et al. 2016). A combination of physical and/or cognitive skills are needed for each of the task execution. In manual operations, human capabilities are used to achieve the required skills. The automation of many of these tasks is possible if (a) the skills are achievable through some technology, (b) automation is reconfigurable in a justifying time, and (c) the technology is safe to be used alongside humans (responsible to take care of rest of the (difficult) tasks).

A collaborative robot is a mechanical device for object manipulation in direct physical contact with humans (Krüger et al. 2009). The concept of a collaborative robot was introduced by Colgate (Peshkin M et al. 1996)(Edward et al. 1996)(Wannasuphoprasit et al. 1998) as an Intelligent Assist Device (IAD) that can manipulate objects in direct collaboration with a human operator. The concept was embraced by robot manufacturers and a newer hardware in the industrial robotics arena was introduced referred to as cobots (collaborative robots). The purpose was the automation of physical tasks in closer and safer proximity to humans.

Today intelligent robots, as capable members of human-robot teams, are being envisioned for homes, hospitals, offices and for more advanced settings such as space exploration (Gao *et al.*, 2021). The technological development in robots is expeditious and robots are being discussed as humans' future teammates (Correia *et al.*, 2019). Examples from modern day robotics are: Robonaut, a humanoid robot to work with human astronauts for maintenance operations in space missions (Hoffman and Breazeal, 2004); BigDog – a humanoid military robot to take part in combat operations (Balakirsky *et al.*, 2010); HANDLE- mobile robot to move boxes in the warehouse; SPOT – a nimble robot to climb stairs and to traverse rough terrains; ATLAS- a robot to demonstrate human-level agility (Kamikawa *et al.*, 2021). Besides agility, modern robots are being enabled to be intelligent of their environment and plan their actions accordingly (Fang *et al.*, 2022). Many studies on robots as human teammates have been reported in the past years covering both the technical and social facets of team forming and performance (Major and Shah, 2020).

This paper, in light of the manufacturing paradigm model presented by Koren (Koren, 2010), presents that a new manufacturing paradigm is possible. A better understanding of it is important to better utilize it, get aligned with it and be prepared for it. The contributions of this paper are:

- The potential reshaping of industrial assembly in terms of human-robot teams
- Framework to develop intelligent reconfigurable and repurposable adaptive assembly (IRRAA)
- The technological enablers and building blocks of IRRAA

## 2. Research background

### a. Manufacturing paradigms

The manufacturing industry, since its birth around two centuries ago (Upadhyaya Upadhyaya *et al.*, 2017), has experienced several revolutionary paradigms driven by (1) the plight of new market and economy, and (2) emerging societal imperatives driven by customers (Koren, 2010). Koren explains that the desire of customization to individuals' tastes, preferences and/or buying power are the morphemes that shape societal needs and market demographics (Yao and Lin, 2016).

The altering market demographics require manufacturers to develop new types of manufacturing systems (to produce products), and new business models (to sell products). The integration of new manufacturing systems with new business models and with new product architecture constitutes new manufacturing paradigms (Mironov *et al.*, 2009) (Koren, 2010).

As explained by Koren, the societal need to reduce automobile price was materialized by the invention of moving assembly line (a new manufacturing system) that was combined with the technology of interchangeable parts. Hence the paradigm of mass production was realized (Figure 1).

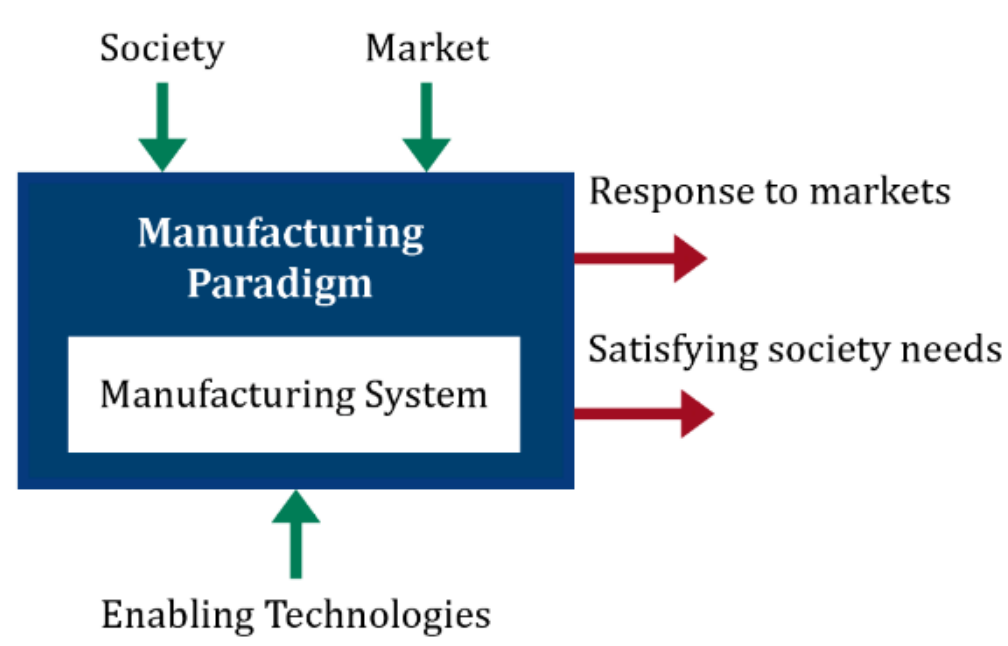

*Figure 1. The birth of new manufacturing paradigms* (Koren, 2010).

## b. Anatomy of industrial assembly

Assembly can be described both as a noun (object) and a verb (action). Assembly, as an action, defines the sequential aggregation of parts, sub-assemblies and/or software (see Figure 2) resulting into functional products (Wiendahl et al. 2015). The parts and sub-assemblies are often manufactured at different times and locations. Assembly tasks emerge from the need to build together all the parts into final product of higher complexity in a required quantity and in a given time period (Nof et al. 2012). Large number of variants, dexterous grasping of components and frequent production changes are a few characteristics that differentiate assembly from other manufacturing processes (Malik, Andersen and Bilberg, 2019). Industrial assembly can be differentiated from a non-repetitive or hobby-assembly by its goals of efficiency, productivity and cost-effectiveness.

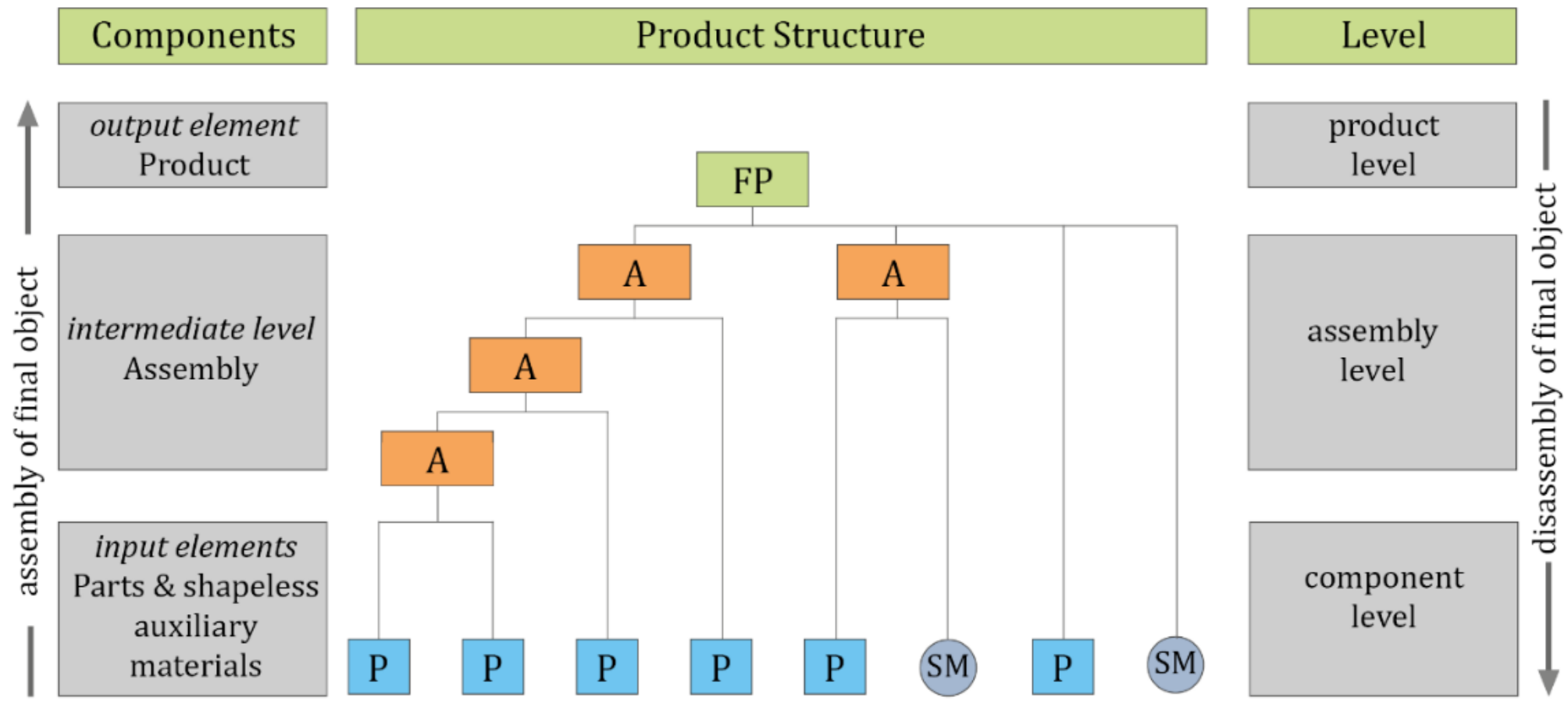

*Figure 2. Components and structure of an assembly product* (Wiendahl, Reichardt and Nyhuis, 2015)

** (FP: finished product; A: assembly; P: part; SM: sub-assembly.*

Since assembly, as a product, is composed of different parts and components (Figure 3), a task, in an assembly process, is defined as the manipulation of human or robotic arm to reach a target and execute an action (e.g. grasp, insert and release) to integrate these parts (Pellegrinelli *et al.*, 2016). It is possible to decompose an assembly task into three components i.e., grasping the object being assembled, transporting the object to a desired location and placement of the object in a required orientation (Andersen *et al.*, 2014). The information content of these tasks is what referred to as assembly complexity (Asadi, Jackson and Fundin, 2016).

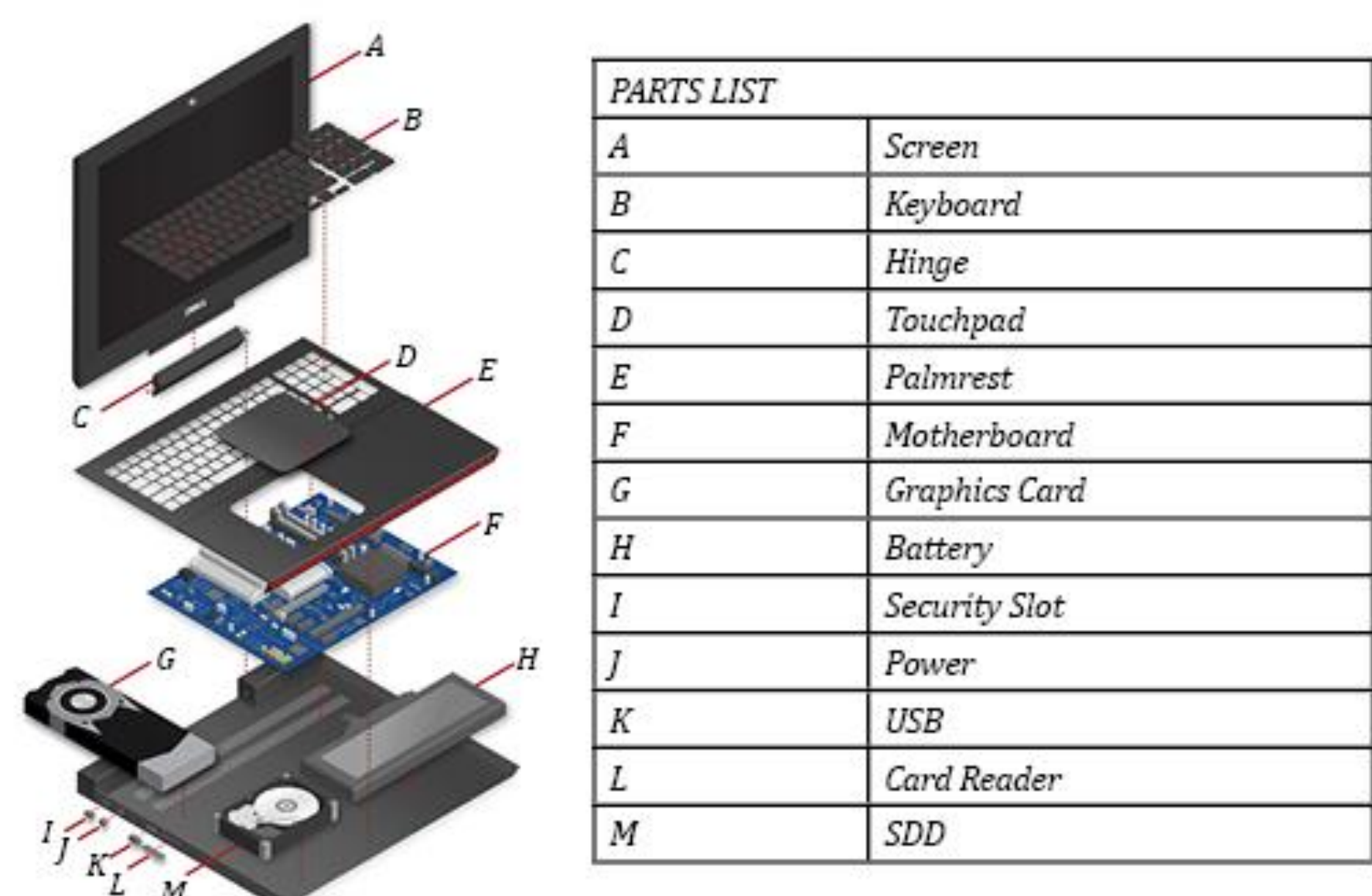

| PARTS LIST | |
|---|---|
| A | Screen |
| B | Keyboard |
| C | Hinge |
| D | Touchpad |
| E | Palmrest |
| F | Motherboard |
| G | Graphics Card |
| H | Battery |
| I | Security Slot |
| J | Power |
| K | USB |
| L | Card Reader |
| M | SDD |

*Figure 3. Parts of a laptop assembly in an exploded view (Dell Technologies).*

### c. Rationalization in assembly automation

Rationalization in assembly automation implies the efforts to improve the quality and reduce cost of the assembled products (Nof et al. 2012). A variety of engineering and management methods can support rationalization such as development of new materials, time-and-motion studies, methods analysis and improvement, new manufacturing and joining techniques, product development and design, mechanization and automation. Figure 4 shows the evolution of assembly starting from pre-industrial craft assembly to assembly lines and to modern day assembly cells.

Conventional heavy load industrial robots, even though being immobile and difficult to reconfigure, are at the backbone of a great proportion of industrial automation (Malik & Bilberg 2017). They are applied in various areas of value addition in a manufacturing value-chain to boost productivity. But when it comes to assembly operations, they are not practical. Assembly accounts for only 7.3% of robotic sales (Shneier et al. 2015).

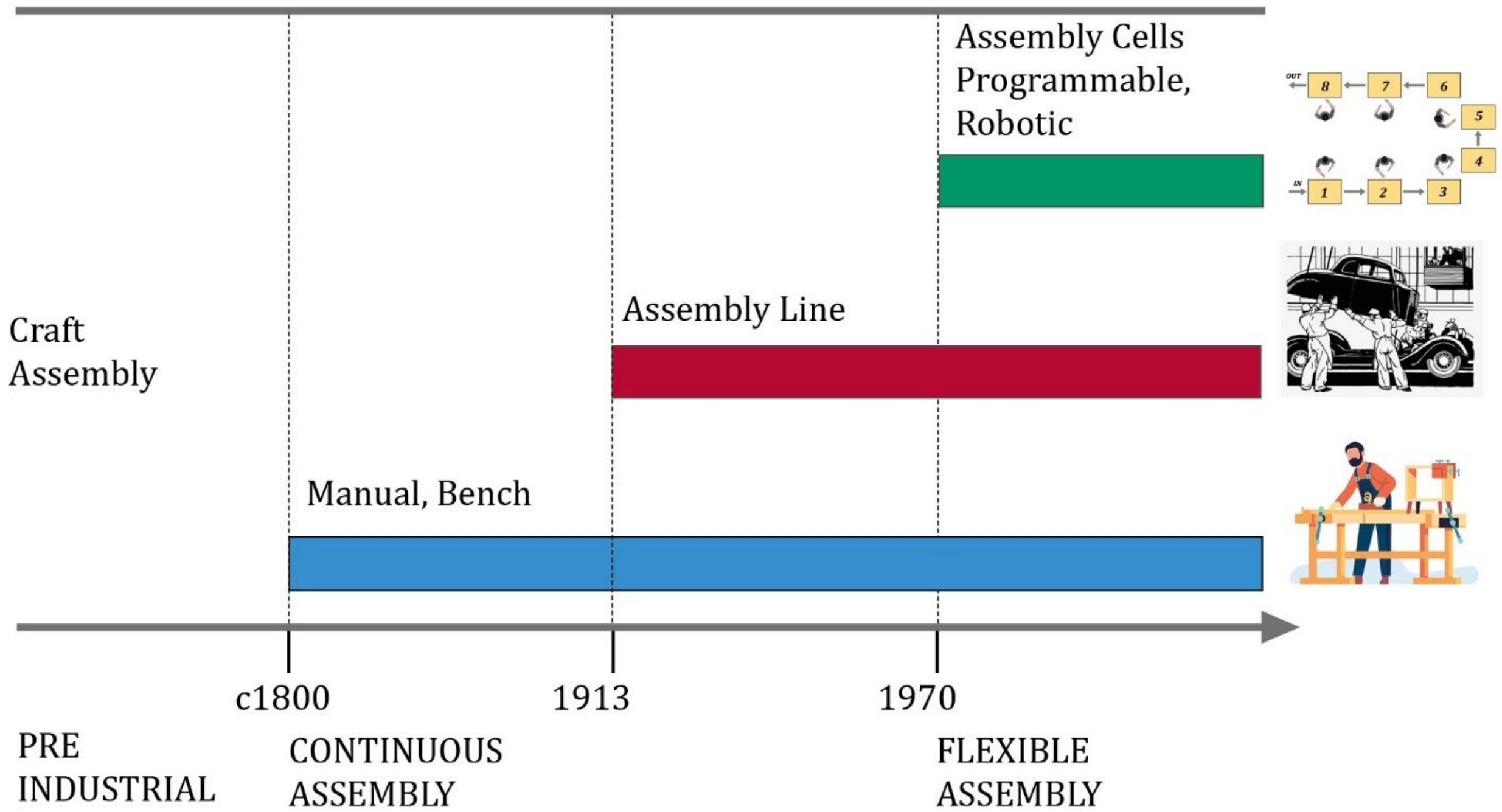

*Figure 4. The evolution of industrial assembly.*

### d. Challenges in assembly rationalization

(Fasth-Berglund & Stahre 2013) described the high degree of intelligence required in the assembly process as a challenge for automation of final assembly systems. Tight tolerances, extensive use of a variety of tools/assistive devices and difficult orientation requirements further add to the assembly complexity for (robotic) automation (Shneier et al. 2015). Therefore, the impact of mainstream automation on assembly is significantly less than other manufacturing domains (Fang *et al.*, 2022). As a result, for manufacturing of discrete products, Boothroyd (Boothroyd 1984) recognized assembly as the most labor intensive process. Interestingly, the scenario has not changed much yet after 38 years (Simões *et al.*, 2022).

The challenges on the way to assembly rationalization are:

- Assembly tasks are highly product specific exhibiting a variety of handling, joining, adjusting and testing
- The continuously shifting market dynamics require assembly, being the final production stage, to adjust with regards to timing, batch sizes and product style

As a result, rationalization measures adapted in medium and short-run manual assembly are limited to making improvements in work-structuring and workstation-design through motion study to avoid wastes. However, a human-centric production system can have drawbacks (Yamazaki et al. 2016) such as:

- Lengthy training process for the operator
- Risk of job-hopping (especially for specialist workers)
- Difficulty to find right trained and skilled operators
- Inconsistency and impossibility of zero defect

### e. Teams of humans and robots for assembly automation

A high level of automation and maximum exclusion of humans seem to be a best choice to gain productivity but it comes with several challenges as described by Bainbridge in her 1983 acclaimed article, 'Ironies of automation' (Bainbridge 1983). Therefore, when moving towards automation, humans, should not and cannot be replaced by machines rather machines must assist humans in increasing their productivity or informing them the state of the system when automation will be nearing its abilities (Norman, 2015). For this purpose, a growing research base is getting available examining human capabilities involved in manual work and the changes observed when interacting with automation (Parasuraman & Wickens 2008).

A human-robot team can be defined as, *'An entity of one or more humans and one or more robots that work together by sharing time and/or space to achieve common goals.'* When developing a human-robot collaborative workspace, the phenomena of teaming between humans and robots can be studied at different dimensions. These includes social, technical and competency. Besides several developments in the use of collaborative robots the question of the level of engagement between human and robot (Fast-Berglund and Stahre, 2013) (Michaelis *et al.*, 2020) is not mature yet. This signifies a lack of measurement science to assess the types and levels of human robot collaboration and the safety provisions. This aspect is important for manufacturers to develop human-robot teams to be easily integrated into operations both for new as well as existing systems. The development of a human-robot team should be much similar and as simple as to develop humans' teams.

Two fundamental properties of a team's members are capability and adaptability. Capability of an entity is defined as the objective-achieving property that helps it realizing its overall mission (Malik and Bilberg, 2019b). For production systems, the capability of a resource is to have the ability to accomplish the desired tasks. Next to capability, the other desired property from a production resources is adaptability. Adaptability is the ability of a resource to switch to new objectives. The natural intelligence of humans enables them to easily adapt to production changes and requirements (Peschl et al. 2012) (while for machines, some kind of hardware/software modifications are needed to make it adapt to new objectives).

Capability classification (Romero et al. 2016) is described as: (a) Physical capabilities: humans' capacities to undertake physical work, characterized with multiple attributes of physical functions (lift, walk, manipulate) and non-physical functions (speed, strength, precision); (b) Sensorial capability: the capacity of an operator to collect and use data from the environment to accomplish the daily tasks; (c)

Cognitive capabilities: operator's ability to perform mental tasks such as reasoning, decision and perception.

Assembly automation has often been described as the mechanization of physical tasks requiring physical capabilities. However, assembly process is a complex process with both physical and cognitive tasks. Physical tasks are the basic tasks such as handling, joining, screwing etc. and cognitive tasks deal with control and support of physical tasks (Frohm et al. 2008).

## 3. State of the art

Efforts to develop human-centered automation rather than machine-centered automation had begun already in 1989 (Hancock & Scallen 1996). Machines are good at most of the tasks where humans are bad at; creating opportunities to devise systems based on the bests of each other (Malik and Bilberg, 2019a). The success and productivity of human-machine collaborative systems depends upon the synchronized orchestration of humans and robots that occurs at various spatio-temporal scales (Christensen, 2016). Carefully designed systems can make humans an integral part of the automation system at a level appropriate to their abilities.

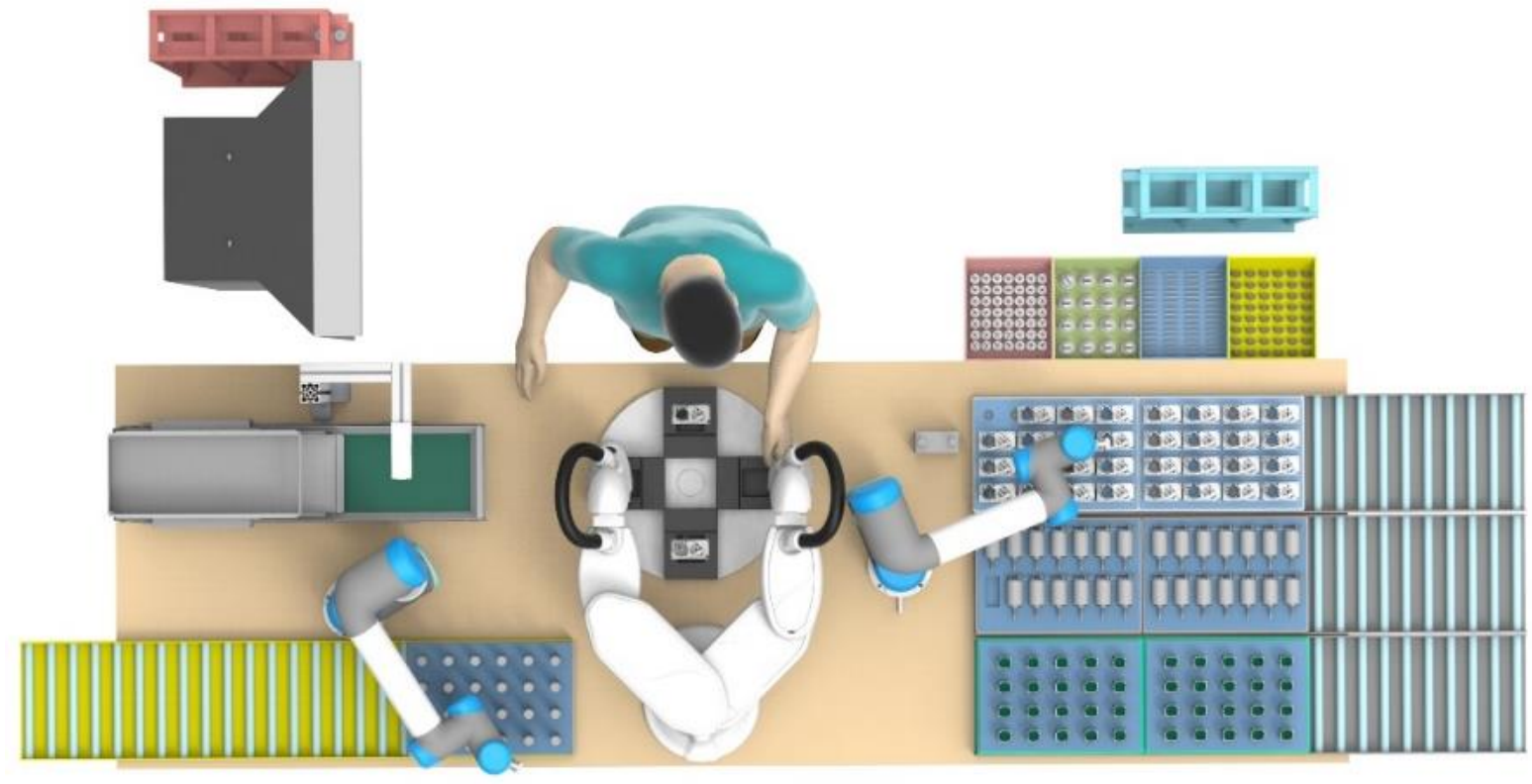

*Figure 5. A schematic of human-robot collaboration for assembly tasks.*

The cobots available today are articulated with multiple rotational joints (degrees of freedom) resulting in a flexibility to reach every coordinate of their workspace in multiple configurations (Djuric et al. 2016). In addition to the safer interaction with humans, cobots through a light-weight structure and easier reconfiguration, can enable mobility and adaptability. These benefits are tempting the manufacturers to automate processes that conventionally were not easy to automate due to closer presence of humans. The benefits are aimed to enhance productivity, efficiency, better ergonomics and improved safety (Wannasuphoprasit et al. 1998)(Grigorescu et al. 2010)(Fast-Berglund et al. 2016).

Several HRC assembly cells have been reported in literature in the past years. A self-learning approach for robots through observation with minimal human effort was presented by (Ji *et al.*, 2021). Use of HRC assembly in wind turbines manufacturing has been reported by (Malik, 2022). (Mohammed and Wang, 2018) reported the usability of a brain-wave controlled HRC assembly system. Programming approaches for dual arm robots for assembly operations have been reported by (Makris *et al.*, 2014). An adaptive framework for human-robot collaboration has been presented by (Buerkle *et al.*, 2022). However, all these approaches discuss one human to one robot (limited) collaboration and the work cell is not an element of a time constrained system.

Other approaches to HRC assembly have also been reported such as assembly balancing. Assembly balancing is a classical problem of assembly systems aimed to evenly distribute tasks among workstations avoiding any idle or waiting time at any station. Assembly balancing becomes even challenging in human-robot collaboration (Fera *et al.*, 2020). A genetic approach for assembly line balancing problem in human-robot collaborative assembly was presented by (Dalle Mura and Dini, 2019). Another approach using digital twin to perform assembly process balancing in HRC has been presented by (Bilberg and Malik, 2019). The assembly balancing in HRC is aimed at minimization of assembly system cost, person hours and on the level of collaboration with robots (Fera *et al.*, 2020).

## 4. IRRAA: Intelligent Reconfigurable & Repurposable Adaptive assembly

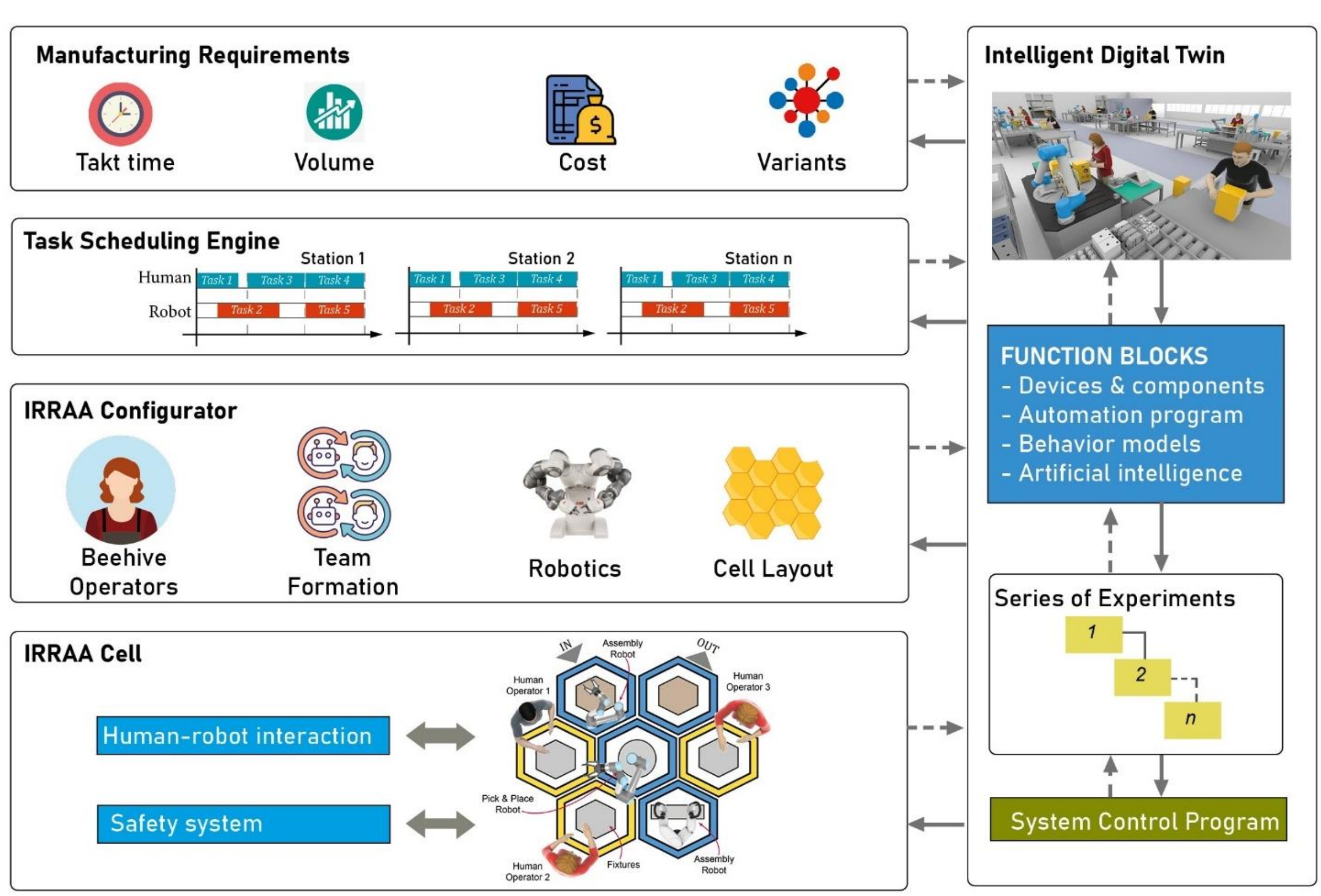

*Figure 6. Framework of IRRAA.*

This section presents a generic framework towards an Intelligent Reconfigurable & Repurposable Adaptive assembly system (Figure 6). Cobots, equipped with required safety devices, can conjoin the strengths and weaknesses of machines with fellow humans, creating a workspace exhibiting the right amount of automation (Müller et al. 2016). Cobots can provide strength and endurance of machines, while human operators offer motive-power, flexibility and intelligence to execute a task. The result is a balanced automation formed at the intersection of flexibility of humans and efficiency of machines. Such a scenario needs to address developments in several dimensions.

## 4.1. Building blocks

The framework consists of a human-robot multi members team in a time sensitive environment. A digital twin of this system is available to design, validate and monitor its operations. The physical system needs to take benefit from innovative hardware solutions. It also has a domain of team configuration. The control engine is the key that examines a process and develops an HRC composition and a layout structure is developed accordingly.

The building blocks are discussed in the below section:

### a. Bee-hives inspired modular architecture

The layout of industrial assembly has evolved from straight assembly lines to S, U, Z, L shaped flexible assembly cells. An assembly cell must have a definite structure for inflow and exit of material. To make best use of human-robot collaboration in a flexible, and reconfigurable structure, innovative approaches are needed. Though flexible U-shaped assembly cells have been fulfilling the needs efficiently but they are limited in capacity adjustment (Gil-Vilda *et al.*, 2017). Additionally, the introduction of various robots will make the layout cumbersome.

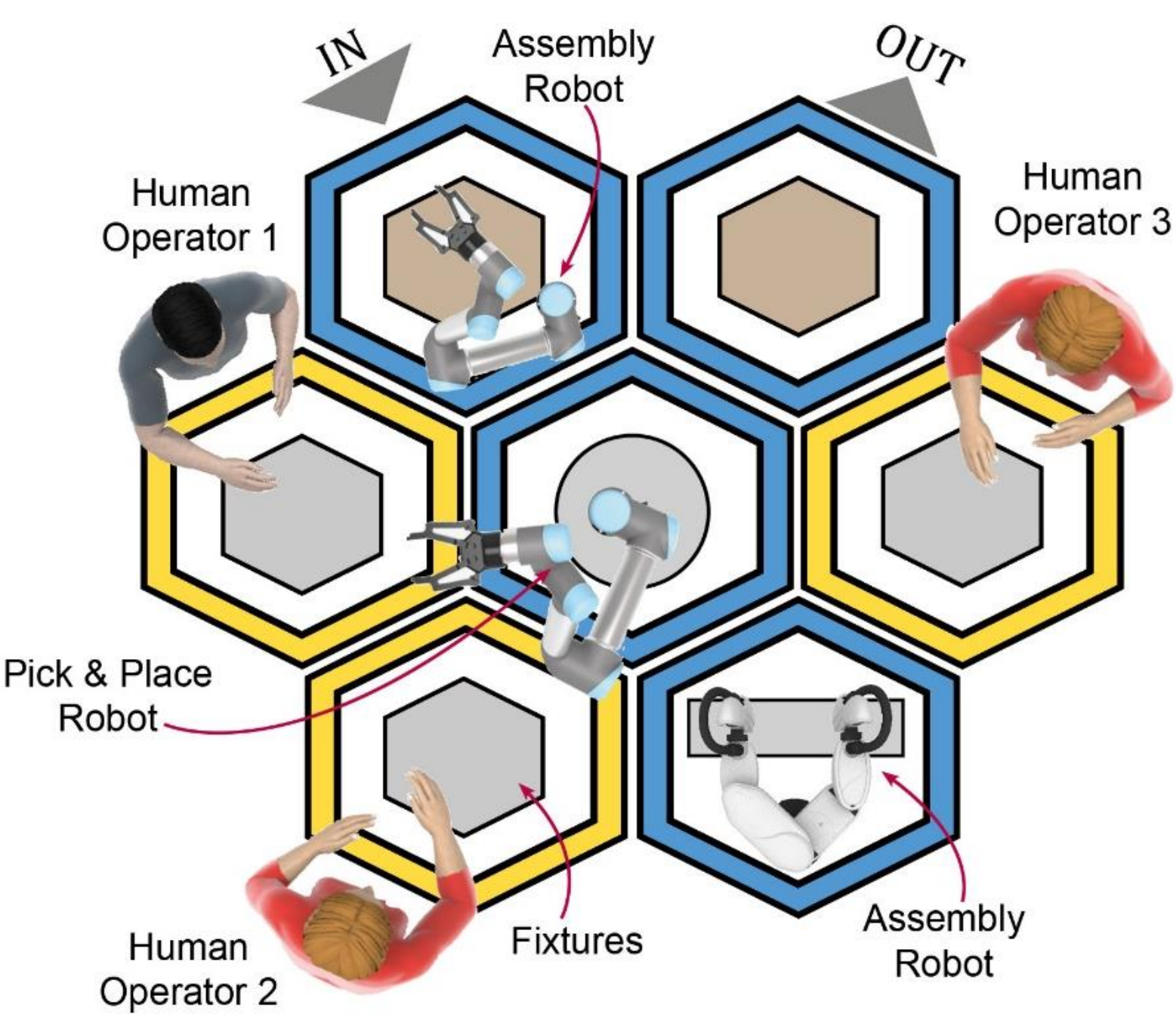

*Figure 7. Architecture model for human-robot collaboration.*

A proposed layout inspired by beehives for IRRAA is shown in the Figure 7. The beehives are hexagonal shapes and offer the possibility to add or remove interchangeable modules. By having such a layout, additional modules can be added or removed as the need arises. Depending upon the task complexity, a layout must be conceivable quickly ensuring less material handling and a high degree of human-robot cooperation.

Modularization is analogous to toys construction bricks and is key enabler to achieve manufacturing reconfigurability (Joergensen, 2013). Though modularization can be enabled in the products being produced [47] [48] and also in its manufacturing system but product modularity is out of scope of this study. Modularization in a manufacturing system is enabled through a universal architecture in the mechanical design of its elements referred to as modules. When needed, the capacity or capability adjustment is enabled by quick and economic inclusion or exclusion of modules supported by its universal architecture [46]. The modularization can also be enabled in the information flow.

Since modularization is an ideal solution for reconfigurability [50] it must be ensured in IRRAA cells that combines automation and robotic skills. Different types of robot manipulators with varying configurations, technical specifications, and degrees of freedom can be used to accomplish a variety of tasks. As per task complexity and assembly sequence, modules can be added in a specific order to form an assembly cell. Such a reconfigurable and modular HRC cell may address demand fluctuations [46]. Modules can be readjusted for task distribution criteria and additional modules maybe added if the demand fluctuates (see Figure 8).

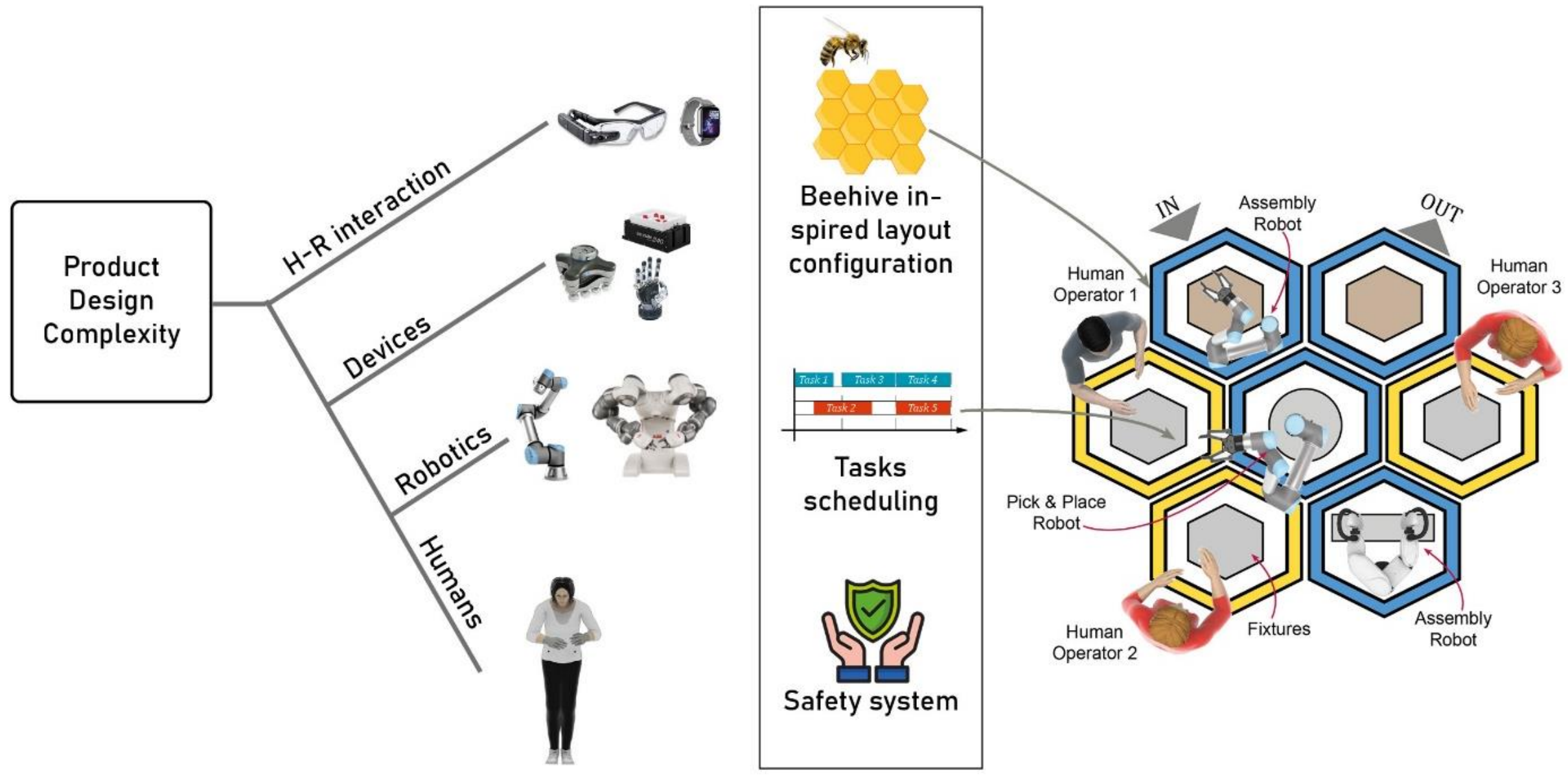

*Figure 8. Hardware configuration model for IRRAA.*

Another important aspect needs to be considered is 'plug and play' hardware for ease of configuration. It suggests to have hardware that is portable and quick to implement. Examples of generally used hardware for cobot systems are grippers, cameras, feeding systems, fixtures, screw drivers etc.

## b. Human-robot team configurations

When working in teams, teammates need to have three capabilities. They need to know what teammates are thinking, anticipate teammate's actions and make fast adjustments as needed (Shah et al. 2011). Though it is a challenging subject in human-robot interaction in unstructured environments. However, assembly cells are semi structured environments where the requirement of robot intelligence is comparatively less as compared to non-manufacturing spaces (such as homes and hospitals). Therefore, robots - as teammates - can be exploited in assembly cells given the safety constraints allow it.

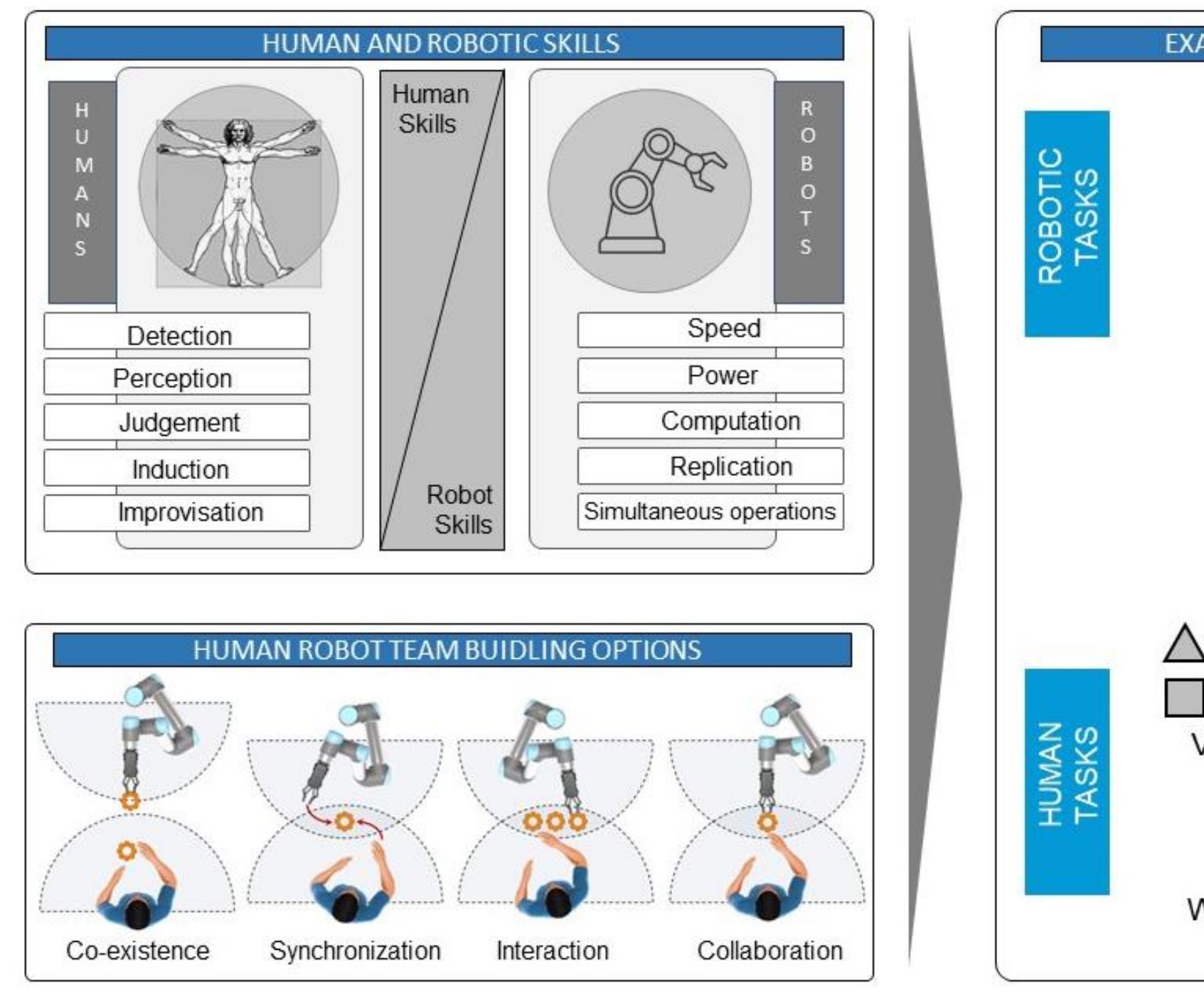

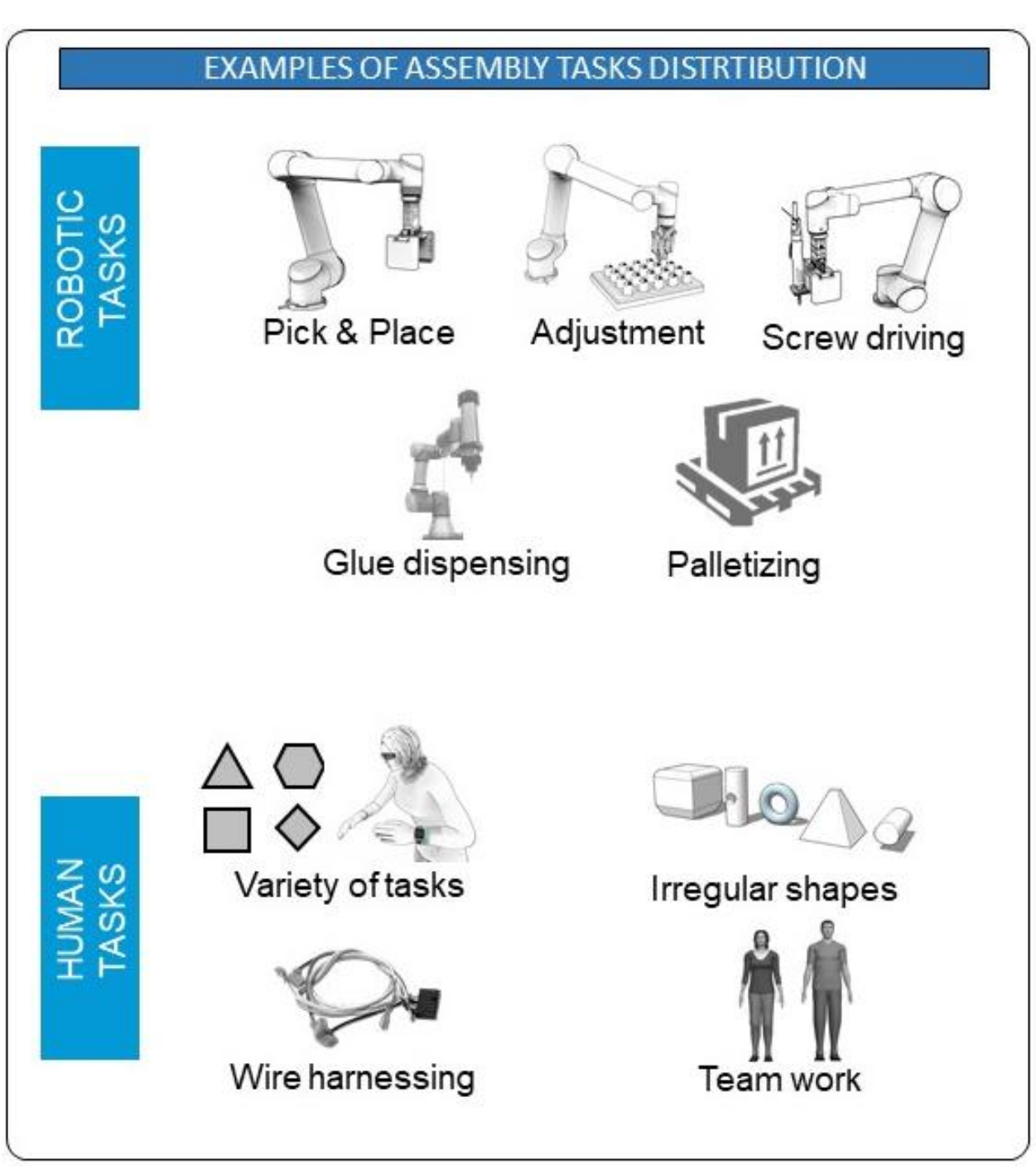

*Figure 9. Architecture model for human-robot collaboration.*

HRC aims a shared work environment where human(s) and robot(s) share their best and complimenting competencies to solve some complex tasks (Wang et al. 2019). The result is a society of agents capable to solve problems that would otherwise be impossible to accomplish under given constraints. The cooperation between the human and robot agents can be at various types of collaboration and team composition (see Figure 9 and 10).

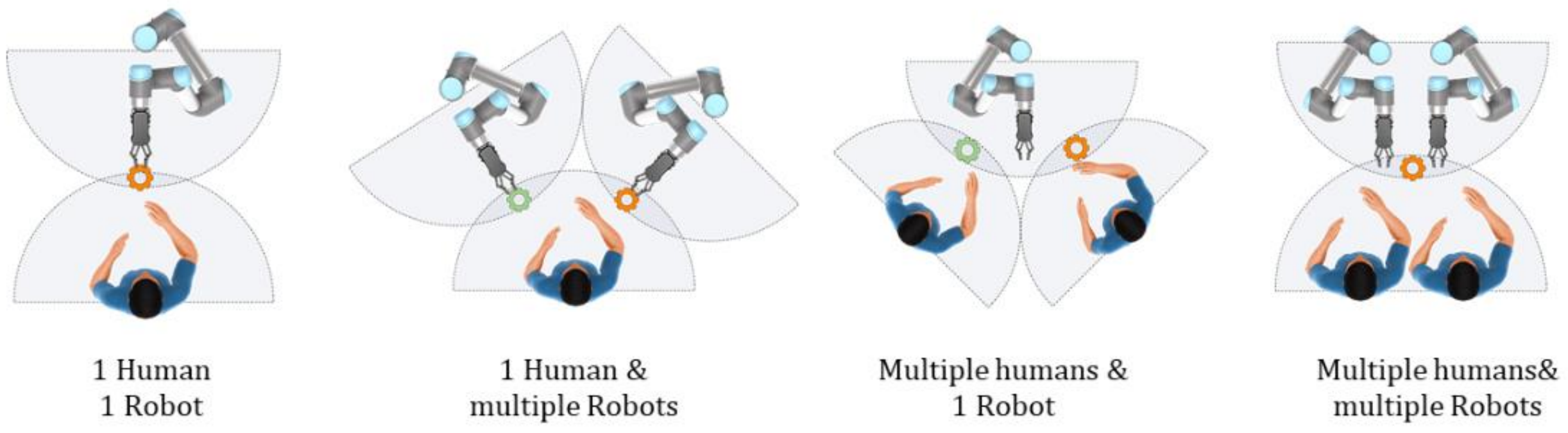

*Figure 10. Model for human-robot teaming.*

The possible types of interaction or team-composition strategies are shown in Fig. 3. The level starts from one human and one robot, to multiple humans and multiple robots. Contradictory to Janco's taxonomy, where team composition is defined at eight levels (Fig), we reduce the types by moving the level of interaction to other axis because modern industrial experimentations have surpassed the possible interactions compared to Janco's work.

An aspect of collaboration in human-robot teams is task cohesion (i.e. commitment to team's goals and tasks). Task cohesion in a team increases team performance. More cohesive teams display greater amount of efficiency, viability, and satisfaction (Vegt et al. 2018). This aspect is discussed more in detail in Chapter 3.

Secondly, the teaming also needs to be studied in its very contexts. Many researchers hypothesized the variation in team performance explained by team structures. The team structure is defined as the team relationships defining the allocation of tasks, responsibilities and authority (Stewart & Barrick 2000). Regardless of the team structure, humans and robots can work good only when they are working in a synergistic fashion. A system that enables humans and robots to communicate and to collaborate is less brittle, more robust and better performing (Fong et al. 2006). Human teaming is a well-researched area and numerous studies are available on human teaming in production settings but little work is available on applying these models to human-robot teams (Nikolaidis & Shah 2012).

### c. Intelligent digital twins

Digital twins (DT)are data connected digital copies of elements and dynamics of a physical system (Malik and Brem, 2021b). A DT consists of a digital space, a physical space and data and information exchange between them (Grieves and Vickers, 2017). Digital twins have been discussed and argued as a solution to design, develop, commission and operate complex systems including manufacturing systems (Mourtzis, Doukas and Bernidaki, 2014) (Erkoyuncu et al., 2020) (Wilhelm, Beinke and Freitag, 2020).

In the IRRAA framework, DT is the brain of the production system throughout its lifecycle. It supports the design activity, commissioning, safety assessment, training and every reconfiguration. The new strategies are simulated, compared with historical happenings and more informed decisions are made and translated to the real system.

New technologies such as Omniverse, Metaverse and A.I. enabled gaming engines have created new opportunities to build high fidelity digital twins, and utilize gaming intelligence methods along with realistic physics behavior. These technologies are useful when programming, optimizing and validating robotic devices.

### d. BeeTeam: assembly operators

Scholtz (Scholtz 2002) defined the roles that humans may take in an HRC scenario. These roles are; a- supervisory role: human supervising the behavior of the robot but not directly controlling it; b- operator: a higher interaction with robot needing to change robot's behavior; c- teammate: human working with a teammate robot to accomplish a group of tasks with divided tasks in between human and robot; d- mechanic: human changing the robot hardware or programming it; e- bystander: when human is not controlling the robot but understands its behavior in order to execute some task within robot space. A dynamic switch between these roles can be observed during performance of a goal.

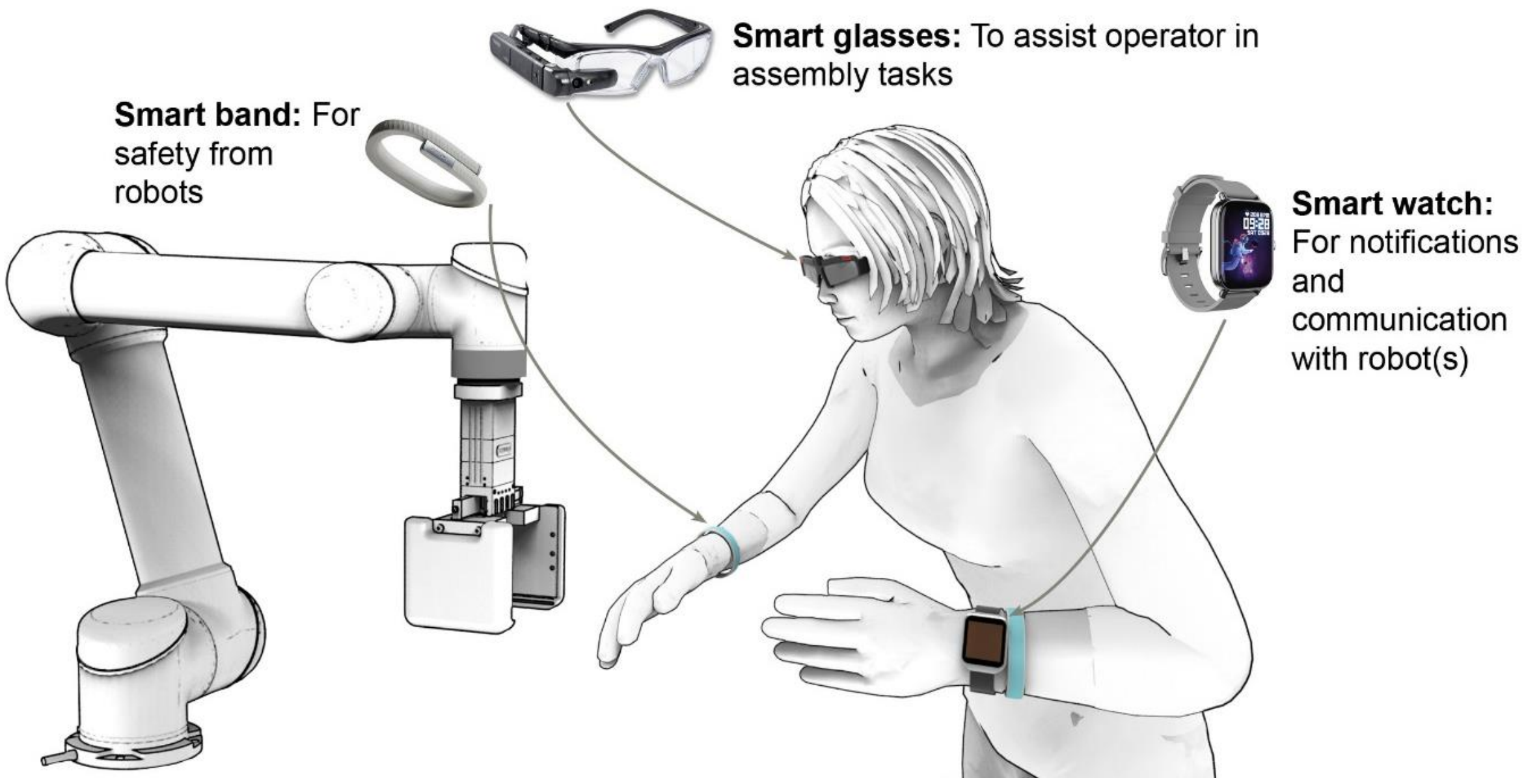

*Figure 11. Smart assembly operators.*

Emerging wearable techn ologies can enable safe and cognitively automated operators (Figure 11).

**Smart watche**s can be used for direct two way communication between human and robot, as well as with MES. Any significant notifications can be passed to the operator using smart watch. Also the operators can transfer a command to the robot controller using smart watch.

**Safety bands**, similar to Vive trackers, can be used by human operators as safety devices. It will add an additional safety layer. The robots will stay away from the band thus ensuring safety to human operators.

**AR glasses** are becoming user friendly. They can be used to communicate assembly instractions, any special guidelines to operators.

## 5. Possible future landscape

Explore the challenges and opportunities

### a. Design for collaborative assembly

Design for Assembly (DFA) is a set of techniques to design products for ease of assembly. Since the assembly is most labour-intensive work and has traditionally been remained away from main-stream automation, the DFA techniques, by its very nature, are based on human abilities of assembling a product. The combination of humans and machines has the potential to increase automation and product mix. A collaborative robot, known as a cobot, is a mechanical device for manipulating objects in human-machine cooperation through direct physical contact. With the right level of adaptability, together with required safety devices, the cobots can enable the right amount of automation, however, they are not as flexible as a human hand. In the context of human-robot collaboration, the required skills for assembly are different from the traditional approaches. It is a merge of design for human-assembly and design for robotic-assembly. This paper documents a set of design rules as a guideline for the future design of cobot assembly (DFCA). With industrial use cases, the DFCA techniques are validated.

### b. Use of artificial intelligence

The development in gaming engines (such as Omniverse) can enable to combine the effects of co-simulations (agent-based simulation, deterministic continuous simulation, and probabilistic discrete simulation), with data from sensors and devices to develop a digital twin in the entirety of the lifecycle of a manufacturing system. A close to reality visualization, large scale resources, and machine learning based behavior can help to probe, sense and respond to any new happening in a complex system. One solution that can come out of this is an adaptive behavior of robots (or AGVs) when interacting with humans and intelligently (online) generating new robot path and making it possible to have zero robot-programming.

## 6. Economic significance of IRRAA cells

For any product produced, over 50% of the total labor cost and time is incurred in assembly (Nof et al. 2012). The statistics differentiate from one sector to other, e.g. the assembly labor costs are 50% for automotive and 20-70% for precision instruments (Nof et al. 2012). Assembly also represents 25-30% of the total cost of any manufacturing company (Shneier et al. 2015) and one-third of the workforce employed in the manufacturing sector (Li et al. 2014) (Malik & Bilberg 2019). The statistics illustrate the significance of assembly for the GDP (gross domestic product) of countries. The three leading industrialized countries (US, Japan and Germany) have their 40% of manufacturing in assembly (Nof et al. 2012). While the total value add of assembly is higher than its percentage of total manufacturing.

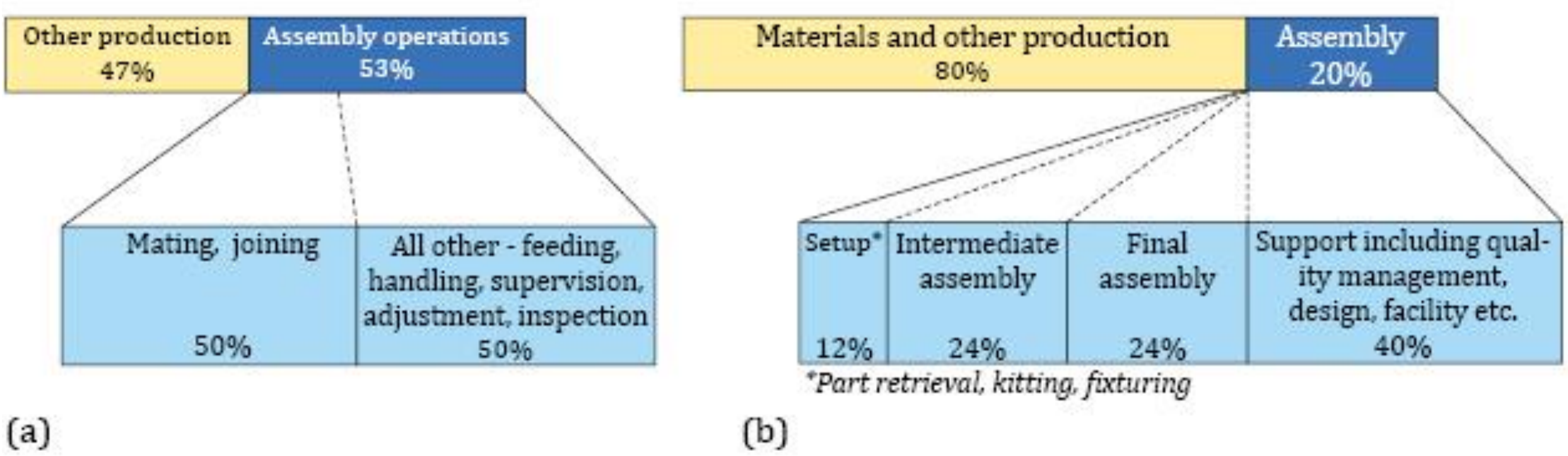

*Figure 12. Typical average breakdown of (a) production time and (b) production cost of industrial products* (Nof et al. 2012).

The statistics reveal the assembly significance and opportunities of potential savings by improving the final assembly systems. Yet, assembly has been the least interesting area for the drive towards automation. A high level of sophistication has been achieved through automation in components manufacturing resulting in a relatively low proportion of labor costs. By contrast, besides the economic and technical significance of assembly, the rationalization measures developed for assembly automation are limited. A study documented that over 90% of assembly tasks are manual (Fast-Berglund et al. 2016), an example of this is smart phone manufacturing. It is not wrong to say that assembly has not been an area of prime research for automation experts.

## 7. Challenges towards IRRAA

The human-robot hybrid work environment is complex and dynamic. Since these systems are supposed to offer high product variety, they need to be able to continuously extend and adapt to various configurations during their operation. Methodologies and frameworks have been presented for developing human-robot teams (Tan et al. 2009) (Francalanza et al. 2014). It is important that such systems are quickly validated during design, development and operation. Safety is another vital consideration in designing HRC workspace (Pedrocchi et al. 2013) (Zanchettin et al. 2016). The workspace should ensure that no harm is caused to human, either directly or indirectly, both during operation or if the system goes into a failure. New approaches are required to design and reconfigure the robots and the production system thus minimizing the required time and effort.

Team is defined as *'entity of two or more interdependent individuals that work together toward common goals'* and teaming is defined as *'the activity of working together as a team'*. To achieve a common goal human, need to have common mental models. But for humans to have a robot colleague having a common mental model is difficult. The traits of human teamwork are:

- Know what teammates are thinking
- Anticipate teammates actions
- Make fast adjustments as needed

## 8. Conclusion

An area of manufacturing that accounts for 50% of the production time, 20% of the production cost and one-third of the human labor is assembly. Yet it has been least interesting for the drive of automation. Collaborative robots have shaped a possibility that humans and robots can work together to share the physical tasks thus freeing up humans from repetitive tasks and getting involved in value added tasks. It can be envisaged now that the future of industrial assembly is in the form of adaptive human-robot-team cells.

It is worth mentioning that besides all the advantages documented about cobots, the challenges of using cobots in industrial environments are still enormous. The puzzling revolves around adaptability, shorter cycle times, and safety. Research community is trying to answer these questions (refereneces). Nevertheless, cobots have flared up another dream of humans and robots working as teams in future factories.

With continuous increase in customers' expectations for customization and personalization, the change, in products and systems, is becoming a manufacturing constant. The proliferation of variety is present in wide range of products (ElMaraghy *et al.*, 2013). To meet with this challenge, change enablers and adaptation mechanisms in manufacturing are increasingly needed. Aligned to this, innovative product designs, manufacturing processes and technologies are needed to remain competitive, flexible and responsive (Koren and Shpitalni, 2010). It is well aligned with the Charles R. Darwin's statement in his book *'On the origin of species':*

*"It is not the strongest species that survive, nor the most intelligent, but the ones most responsive to change"*

To address the notion of being resilient, new generation of assembly systems need to be adaptable, flexible and automated. A possible new paradigm can emerge that promises mass customization. Assembly has the highest potential to enable product variety thus mass customization can be the most probable outcome of an assembly system that is optimally automated yet being flexible.